\documentclass[twocolumn,a4paper,10pt]{article}
\usepackage{geometry}
\usepackage{amsmath,amssymb}
\usepackage{bm,braket}
\usepackage{multicol}
\usepackage{comment}
\usepackage[dvipdfmx]{graphicx}
\usepackage{cite}
\usepackage{authblk}
\usepackage{nidanfloat}

\title{{\Large Critical phenomena around the SU(3) symmetric tri-critical point of a spin-1 chain}}

\author{Tohru Mashiko and Kiyohide Nomura}
\affil{Department of Physics, Kyushu University, Fukuoka 819-0395, Japan}
\date{}

\begin{document}
\small
\twocolumn[
\maketitle
\begin{quotation}
We investigate critical phenomena of a spin-1 chain in the vicinity of the SU(3) symmetric critical point, which we already specified in the previous study.
We numerically diagonalize a Hamiltonian combining the bilinear-biquadratic (BLBQ) Hamiltonian with the trimer Hamiltonian.
We then discuss the numerical results based on the conformal field theory (CFT) and the renormalization group.
As a result, we firstly verify that the critical point found in our previous study is the tri-critical point among the Haldane phase, trimer phase, and the trimer liquid (TL) phase.
Secondly, with regard to the TL--trimer transition and TL--Haldane transition, we find that the critical phenomena around this tri-critical point belong to the Berezinskii--Kosterlitz--Thouless (BKT)-like universality class.
We thirdly find the boundary between the Haldane phase and the trimer phase, which is illustrated by the massive self-dual sine-Gordon (SDSG) model.
\end{quotation}
]

\section{Introduction}
\label{sec:intro}
In quantum many-body physics, phase transitions and critical phenomena are interesting topics.
In low dimensional quantum systems with continuous symmetry, quantum fluctuation is so strong that it prevent the long-range order of the system, which makes critical phenomena complicated.
Elucidating properties of the critical phenomena brings some basic theories to experiments of ultracold atomic systems in an optical lattice \cite{desalvo, gorshkov, taie}.
Especially in the case of the SU(3) symmetric spin-1 chains, there have been several studies \cite{uimin,lai,sutherland,kulish} which investigated the massless trimer liquid (TL) state.
On the other hand, several researchers made proposals of the SU(3) trimer model \cite{schm,solyom,greiter}, as the generalization of the SU(2) Majumdar-Ghosh model \cite{majumdar} of the spin-1/2 dimer ground state.
In the previous study \cite{mashiko2}, the authors investigated the TL--trimer phase transition of the SU(3) symmetric spin-1 chain to specify the critical point.
But there remain unsolved problems about the non-SU(3) symmetric case, in which we add the SU(2) perturbation to the SU(3) symmetric spin-1 chain.
Thus, in this paper, we numerically investigate the critical phenomena by expanding our previous study \cite{mashiko2} to the non-SU(3) symmetric case, based on the theory of Itoi and Kato \cite{itoi}.
We research it by combining two Hamiltonians, the bilinear-biquadratic (BLBQ) Hamiltonians and the trimer Hamiltonians as described later.

We here review a well-known spin-1 model, BLBQ model, whose Hamiltonian is defined with the parameter $\theta$ as
\begin{eqnarray}
\hat{H}_{\mathrm{BLBQ}} = \sum^{N}_{i=1} \left[ \cos\theta \hat{\bm{S}_{i}} \cdot \hat{\bm{S}_{j}} + \sin\theta \left(\hat{\bm{S}_{i}} \cdot \hat{\bm{S}_{j}} \right)^{2} \right], \label{delblbq}
\end{eqnarray}
where $\hat{\bm{S}_{i}}$ is the spin-1 operator at site $i$, $j\equiv i+1$.
The region $-\pi/4<\theta<\pi/4$ is the Haldane phase \cite{haldane1,haldane2}. 
This phase is translationally invariant and massive \cite{nightingale,nomura3}. 
The region $\pi/4<\theta<\pi/2$ is the massless trimer liquid (TL) phase without long-range order.
The point $\theta=\pi/4$, which is SU(3) symmetric, is known as the Uimin--Lai--Sutherland (ULS) point \cite{sutherland,uimin,lai,kulish}, which is exactly solvable with the Bethe ansatz. 
The system at the ULS point is critical, whose universality class is described by the level-1 SU$(3)$ Wess--Zumino--Witten [SU$(3)_{1}$ WZW] model \cite{wess,witten1,witten2}. 
Around the ULS point, numerical studies were carried out \cite{fath,lauch} to determine the universality class by calculating the central charge $c$ and the scaling dimension $x$. 
Itoi and Kato analyzed \cite{itoi} systems around the ULS point with the renormalization group (RG) by mapping the ULS model to the general SU$(3)_{1}$ WZW model. 
They found \cite{itoi} that the phase transition at the ULS point belongs to the Berezinskii--Kosterlitz--Thouless (BKT)-like universality class.
In the BKT-like transition, the system is characterized by the quasi long-range order described by the correlation functions \cite{itoi,stou,lauch,schm}
\begin{eqnarray}
	\left \langle \hat{\bm{S}}_{i} \cdot \hat{\bm{S}}_{i+r}  \right \rangle \propto \cos \left(\frac{2\pi}{3} r \right) r^{-2x} \left(\ln r \right)^{\sigma}, \label{sqcorre-in}
\end{eqnarray}
where $2x$ and $\sigma$ are critical exponents.
Due to the logarithmic correction in Eq. \eqref{sqcorre-in}, it has been difficult to calculate the scaling dimension.
We calculate them by removing logarithmic corrections utilizing the theory of Itoi and Kato \cite{itoi}, which is the expansion of the level spectroscopy \cite{nomura4}.

Next, the trimer states $\left| \psi_{\mathrm{T}} \right \rangle$ with the long-range order on a spin-1 chain were proposed \cite{nomura1,xian}, which is given
\begin{eqnarray}
        \left| \psi_{\mathrm{T}} \right \rangle = 
        \begin{cases}
                \{ \circ \circ \circ \} \{ \circ \circ \circ \} \cdots \{ \circ \circ \circ \} , \vspace{0.5em} \\
                \circ \{ \circ \circ \circ \} \cdots \{ \circ \circ \circ \} \circ \circ , \vspace{0.5em} \\
                \circ \circ \{ \circ \circ \circ \} \cdots \{ \circ \circ \circ \} \circ , \\
        \end{cases}
        \label{trimer}
\end{eqnarray}
under the periodic boundary conditions (PBCs).
Here $\{ \circ \circ \circ \}$ is the singlet state of the three adjacent spins (trimer).
The state of the trimer is given as
\begin{eqnarray}
        &&\hspace{-1.5em} \{ \circ \circ \circ \} \equiv \frac{1}{\sqrt{6}} \left( |1,0,-1 \rangle + |0,-1,1 \rangle + |-1,1,0 \rangle \right. \notag \\
        &&\left. \hspace{4.0em} -|1,-1,0 \rangle - |0,1,-1 \rangle - |-1,0,1 \rangle \right), \label{trimer2}
\end{eqnarray}
where $1$, $0$, $-1$ are spin magnetic quantum numbers, $S^{z}$.
Hamiltonians whose ground state is the pure trimer state were proposed \cite{schm,solyom,greiter}.
In the trimer phase, there exists an excitation gap \cite{greiter} between the threefold degenerate ground state energy and elementary excitation spectrum.
Also, the long-range order of the trimer state is illustrated by the correlation function \cite{itoi,schm} of the trimer order parameter $\hat{T}_{i}$ (see Sec. \ref{subsec:corre}).

\begin{figure}[t]
 \begin{center}
  \includegraphics[keepaspectratio,scale=0.0555]{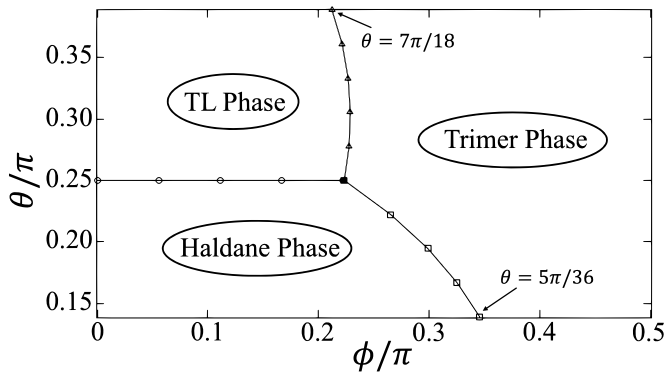}
 \end{center}
        \caption{Phase diagram in the vicinity of the tri-critical point $(\phi,\theta)=(0.223365\pi,\pi/4)$.}
        \label{fig:phase}
\end{figure}

In this study, we numerically diagonalize the Hamiltonian combining the BLBQ Hamiltonian with the trimer Hamiltonian defined as 
\begin{eqnarray}
        \hat{H} \equiv \cos \phi \hat{H}_{\mathrm{BLBQ}} + \sin \phi \hat{H}_{\mathrm{trimer}}, \label{model}
\end{eqnarray}
by changing two parameters $\phi$ and $\theta$, under PBCs to investigate the critical phenomena in the vicinity of the SU(3) symmetric critical point \cite{mashiko2}.
As a result, we obtain the phase diagram (Fig. \ref{fig:phase}) which shows that the the SU(3) symmetric critical point is the tri-critical point among three phases, the TL phase, the Haldane phase, and the trimer phase.
In later sections, we describe the details led up to the phase diagram.
It is the first achievement to numerically calculate phase boundaries of Haldane--trimer transition and TL--trimer transition in the non-SU(3) symmetric cases.
We write the details of the model Eq. \eqref{model}, especially focusing on the definition of the trimer model Hamiltonian $\hat{H}_{\mathrm{trimer}}$ in Sec. \ref{sec:model}.
Section \ref{sec:ren} shows the discussion on the basis of the conformal field theory (CFT) and the renormalization group (RG) by reviewing the paper of Itoi and Kato \cite{itoi}.
In Sec. \ref{sec:tran}, our numerical results of the phase transitions are shown by comparing the theory in Sec. \ref{sec:ren}.
Conclusion and discussion are shown in Sec. \ref{sec:con}.

\section{MODEL}
\label{sec:model}
In order to investigate critical phenomena in the vicinity of the SU(3) symmetric critical point \cite{mashiko2}, we calculate low energy eigenvalues of the Hamiltonian Eq. \eqref{model}.
Here, the parameters $\phi$ and $\theta$ are in the range of $0 \le \phi, \theta \le \pi/2$, and $\hat{H}_{\mathrm{trimer}}$ will be defined in the next paragraph. 

We introduce the Hamiltonian of the trimer ground state by reviewing the paper of Greiter and Rachel \cite{greiter}.
We firstly let $\hat{c}^{\dagger}_{i\gamma}$ ($\hat{c}_{i\gamma}$) denote an operator which creates (annihilates) a fermion with $S^{z}=\gamma=-1,0,1$ at site $i$, and then define the operators $\hat{J}^{a}_{i}$ as
\begin{eqnarray}
	&& \hspace{-2em} \hat{J}^{a}_{i} \equiv \frac{1}{2} \sum_{\gamma,\gamma^{\prime}=-1,0,1} \hat{c}^{\dagger}_{i\gamma} \Lambda^{a}_{\gamma\gamma^{\prime}} \hat{c}_{i\gamma^{\prime}}, \hspace{1em} a=1,\cdots,8 \label{gen} \\
	&& \hspace{-2em} \hat{\bm{J}}_{i} \equiv \left(\hat{J}^{1}_{i}, \hat{J}^{2}_{i}, \hat{J}^{3}_{i}, \hat{J}^{4}_{i}, \hat{J}^{5}_{i}, \hat{J}^{6}_{i}, \hat{J}^{7}_{i}, \hat{J}^{8}_{i} \right)^{\mathrm{T}}, \label{genvec}
\end{eqnarray}
where the $\Lambda^{a}$ are the Gell-Mann matrices.
The operators  Eq. \eqref{gen} satisfy the commutation relations
\begin{eqnarray}
	\left[ \hat{J}^{a}_{i},\hat{J}^{b}_{i^{\prime}} \right] = \delta_{ii^{\prime}} f^{abc} \hat{J}^{c}_{i}, \hspace{2em} a,b,c=1,\cdots,8 
\end{eqnarray}
where $f^{abc}$ are the structure constants of SU(3), and we utilize the Einstein summation convention.
Then, we define $\hat{\bm{J}}^{(\nu)}_{i}$ for neighboring sites $i, \cdots, i+\nu-1$ ($\nu$: integer) as,
\begin{eqnarray}
	\hat{\bm{J}}^{(\nu)}_{i} \equiv \sum_{i^{\prime}=i}^{i+\nu-1} \hat{\bm{J}}_{i^{\prime}}.
\end{eqnarray}
With regard to the trimer states Eq. \eqref{trimer2} on the four neighboring sites, there are only two possible situation as 
\begin{eqnarray}
	\begin{cases}
	\{ \circ \circ \circ\} \circ, \\
	\{ \circ \circ \} \{ \circ \circ \},
	\end{cases}
	\label{auxi}
\end{eqnarray}
where $\{ \circ \circ \}$ is the triplet state of the two adjacent spins, whose state vectors are defined as
\begin{eqnarray}
	\{ \circ \circ \} = 
	\begin{cases}
		\displaystyle{\frac{1}{\sqrt{2}}} (|1,0 \rangle - |0,1 \rangle), \vspace{0.5em} \\
		\displaystyle{\frac{1}{\sqrt{2}}} (|1,-1 \rangle - |-1,1 \rangle), \vspace{0.5em} \\
		\displaystyle{\frac{1}{\sqrt{2}}} (|0,-1 \rangle - |-1,0 \rangle).
	\end{cases}
	\label{triplet}
\end{eqnarray}
The eigenvalues of the quadratic Casimir operator for these two representations Eq. \eqref{auxi} are 4/3 and 10/3 respectively \cite{greiter}.
Therefore, we obtain the auxiliary operators as
\begin{eqnarray}
	\hat{H}_{i} = \left[\left(\hat{\bm{J}}^{(4)}_{i} \right)^{2} - \frac{4}{3} \right] \left[\left(\hat{\bm{J}}^{(4)}_{i} \right)^{2} - \frac{10}{3} \right]. \label{opetri}
\end{eqnarray}
The trimer Hamiltonian is
\begin{eqnarray}
	\hat{H}_{\mathrm{trimer}} \equiv \sum_{i=1}^{N} \hat{H}_{i}. \label{hamitri}
\end{eqnarray}
Furthermore, utilising the exchange operator $\hat{P}_{ii^{\prime}}$, which swaps spins at site $i$ with that at site $i^{\prime}$, there is an equation \cite{greiter}
\begin{eqnarray}
	\hat{\bm{J}}_{i} \hat{\bm{J}}_{i^{\prime}} = 
	\begin{cases}
	\displaystyle{\frac{4}{3}}, \hspace{7.2em} (i=i^{\prime}) \vspace{0.5em} \\
	\displaystyle{\frac{1}{2}} \left( \hat{P}_{ii^{\prime}} - \displaystyle{\frac{1}{3}} \right). \hspace{2em} (i \ne i^{\prime})
	\end{cases}
	\label{gen2}
\end{eqnarray}
From Eqs. \eqref{opetri} -- \eqref{gen2}, we obtain the trimer Hamiltonian as
\begin{eqnarray}
	\hat{H}_{\mathrm{trimer}} &=& \sum_{i=1}^{N} \left[ \hat{P}_{ij} + \frac{2}{3}\hat{P}_{ik} + \frac{1}{3}\hat{P}_{il} + \hat{P}_{ij}\hat{P}_{ik} + \hat{P}_{ik}\hat{P}_{ij} \right. \notag \\
	&& \hspace{2.5em} + \frac{1}{2} \left( \hat{P}_{ij}\hat{P}_{il} + \hat{P}_{il}\hat{P}_{ij} +\hat{P}_{ik}\hat{P}_{il} + \hat{P}_{il}\hat{P}_{ik} \right) \notag \\
	&& \hspace{2.5em} \left. + \frac{1}{3} \left( \hat{P}_{ij}\hat{P}_{kl} + \hat{P}_{ik}\hat{P}_{jl} +\hat{P}_{il}\hat{P}_{jk} \right) \right], \label{trimerp}
\end{eqnarray}
where we define $j \equiv i+1$, $k \equiv i+2$, $l \equiv i+3$.

\section{Renormalization discussion on the SU$(3)_{1}$ WZW model}
\label{sec:ren}
In this section, we review the analytic calculation of the perturbative RG by Itoi and Kato \cite{itoi} to discuss the critical phenomena around the fixed point described by the level-1 SU(3) Wess--Zumino--Witten [SU$(3)_{1}$ WZW] model \cite{wess,witten1,witten2}.

\begin{figure}[t]
 \begin{center}
  \includegraphics[keepaspectratio,scale=0.09]{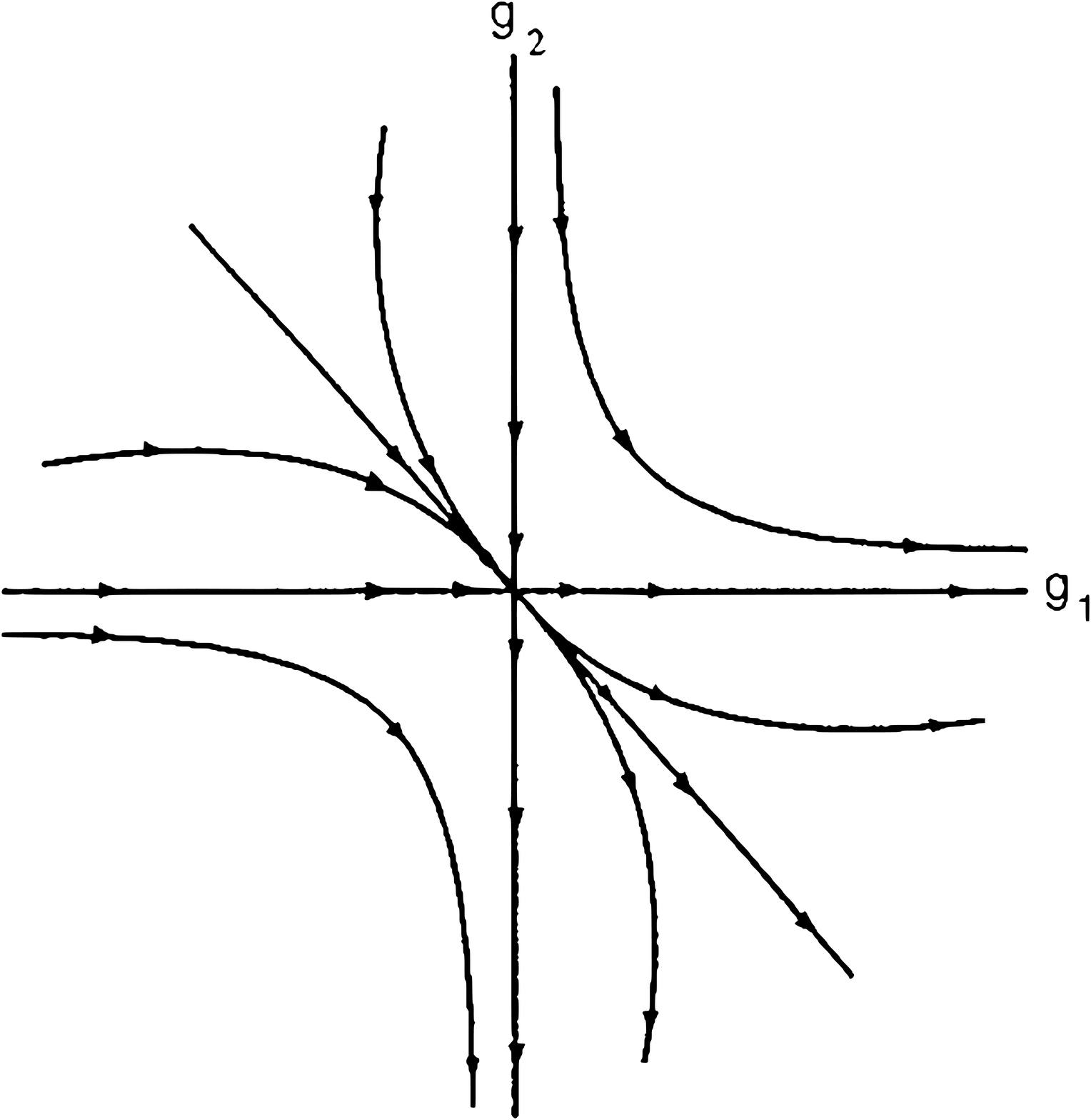}
 \end{center}
	\caption{Trajectory gained by solving, Eqs. \eqref{rg3} and \eqref{rg4}. This figure is taken from Fig. 1 of Ref. \cite{itoi}.}
        \label{fig:flow}
\end{figure}

First of all, we let $x_{0}$ and $x_{1}$ be the time and the position of the field respectively.
We then define $z$ and $\bar{z}$ as
\begin{eqnarray}
z \equiv x_{0} + i x_{1}, \,\,\,\,\,\, \bar{z} \equiv x_{0} - i x_{1}.
\end{eqnarray}
We define the action $\hat{\mathcal{A}}$ as
\begin{eqnarray}
	\hat{\mathcal{A}} \equiv \hat{\mathcal{A}}_{\mathrm{SU}(3)_{1}} + \sum_{i=1}^{2} g_{i} \int \frac{d^{2}z}{2\pi} \hat{\Phi}^{(i)} \left( z , \bar{z} \right). \label{asu}
\end{eqnarray}
Here $\hat{\mathcal{A}}_{\mathrm{SU}(3)_{1}}$ is the action of the free fields in the SU$(3)_{1}$ WZW model \cite{wess,witten1,witten2}.
Both $\hat{\Phi}^{(1)}$ and $\hat{\Phi}^{(2)}$ are operators of the marginal or relevant field with rotational symmetry and chiral $\mathbb{Z}_{3}$ symmetry.
In particular, $\hat{\Phi}^{(1)}$ is SU(3) symmetric, and $\hat{\Phi}^{(2)}$ is SO(3) symmetric.
The scaling variables $g_{1}$ and $g_{2}$ are perturbational parameters, which are both functions of $\phi$ and $\theta$ in Eq. \eqref{model} respectively.
In the vicinity of the fixed point, $g_{1}$ and $g_{2}$ linearly depends on $\phi$ and $\theta$ respectively.
If $g_{2}=0$, the system remains SU$(3)$ symmetric regardless of the value of $g_{1}$.
If $g_{2} \ne 0$, the SU$(3)$ symmetry of the system is broken.
According to Itoi and Kato\cite{itoi}, the renormalization-group equations are
\begin{eqnarray}
	\frac{d g_{1}(l)}{dl} &=& \frac{1}{2\sqrt{2}} \left[3 g_{1}^{2}(l) + 2g_{1}(l)g_{2}(l) \right], \label{rg3} \\
	\frac{d g_{2}(l)}{dl} &=& - \frac{1}{2\sqrt{2}} \left[3 g_{2}^{2}(l) + 2g_{1}(l)g_{2}(l) \right], \label{rg4} 
\end{eqnarray}
where $l \equiv \ln N$.
By solving, Eqs. \eqref{rg3} and \eqref{rg4}, Itoi and Kato obtained \cite{itoi} the flows shown in Fig. \ref{fig:flow}.

Secondly, we review the finite-size scaling based on the RG of marginal operator.
For later discussions, we define the excitation energy as
\begin{eqnarray}
	\Delta E_{S_{T}}(q) \equiv E_{S_{T}}(q) - E_{g}, \label{defdestq}
\end{eqnarray}
where $q$ is the wave number, $S_{T}$ is the total of the spin quantum number $S$, and $E_{S_{T}}(q)$ is the lowest energy at $q$ and $S_{T}$.
Under the finite-size scaling on an infinitely long strip with finite width \cite{cardy1,affleck,cardybook}, we obtain
\begin{eqnarray}
	\hspace{-1.7em}\Delta E_{S_{T}} \left(\pm\frac{2\pi}{3} \right) &=& v_{0} \xi^{-1}_{S_{T}} \left(g_{1}(0),g_{2}(0),N^{-1} \right) \notag \\
	&=& v_{0} e^{-l^{\prime}} \xi^{-1}_{S_{T}} \left(g_{1}(l^{\prime}),g_{2}(l^{\prime}),N^{-1} e^{l^{\prime}} \right), \label{dexi}
\end{eqnarray}
where $\xi_{S_{T}}$ is the correlation length at $S_{T}$, and $v_{0}$ is the spin wave velocity.
Choosing $l^{\prime} = l \equiv \ln N$, we can rewrite Eq. \eqref{dexi} as
\begin{eqnarray}
	\Delta E_{S_{T}} \left(\pm\frac{2\pi}{3} \right) &=& v_{0} \xi^{-1}_{S_{T}} \left(g_{1}(0),g_{2}(0),N^{-1} \right) \notag \\
	&=& \frac{v_{0}}{N} \xi^{-1}_{S_{T}} \left(g_{1}(l),g_{2}(l),1 \right) \notag \\ 
	&\equiv& \frac{v_{0}}{N} \Phi_{S_{T}} \left(g_{1}(l),g_{2}(l) \right),
\end{eqnarray}
where $\Phi_{S_{T}} \left(g_{1}(l),g_{2}(l) \right)$ is a universal function.
By expanding $\Phi_{S_{T}} \left(g_{1}(l),g_{2}(l) \right)$ in a Taylor series up to $O\left(g_{1}(l),g_{2}(l) \right)$, we obtain \cite{cardy1,cardy2}
\begin{eqnarray}
	\Delta E_{S_{T}} \left(\pm\frac{2\pi}{3} \right) \approx \frac{2\pi v_{0}}{N} \left[ x_{S_{T}} + A_{S_{T}}g_{1}(l) + B_{S_{T}}g_{2}(l) \right], \label{destg1g2} \notag \\
\end{eqnarray}
where $x_{S_{T}}$ is the scaling dimension at $S_{T}$.
In the SU$(3)_{1}$ WZW model, the scaling dimension is $x=2/3$.
The coefficients $A_{S_{T}}$ and $B_{S_{T}}$ will be determined in the later subsections.

\subsection{The case of $g_{2}(l)=0$}
\label{subsec:g20}
In the case of $g_{2}(l)=0$, the system is SU(3) symmetric as described previously, and Eqs. \eqref{rg3} and \eqref{rg4} are reduced to
\begin{eqnarray}
	\frac{d g_{1}(l)}{dl} &=& \frac{3}{2\sqrt{2}} g_{1}^{2}(l), \label{rg3a}
\end{eqnarray}
whose solution is
\begin{eqnarray}
	g_{1}(l) = g_{1}(0) \left[1- \frac{3}{2\sqrt{2}} g_{1}(0)l \right]^{-1}. \label{solg1su3}
\end{eqnarray}
As shown in Fig. \ref{fig:flow}, if $g_{1}(0)<0$, the flow is absorbed into the fixed point corresponding to the SU$(3)_{1}$ WZW model.
On the other hand, if $g_{1}(0)>0$, the flow diverges as $g_{1}\rightarrow\infty$, which illustrates the trimer long-range order \cite{greiter,nomura1,xian}.

Also, Eq. \eqref{destg1g2} can be rewritten as
\begin{eqnarray}
        \Delta E_{S_{T}} \left(\pm\frac{2\pi}{3} \right) \approx \frac{2\pi v_{0}}{N} \left[ x_{S_{T}} + A_{S_{T}}g_{1}(l) \right]. \label{destg1g2a}
\end{eqnarray}
By solving equations Eq. \eqref{rg3a}, Itoi and Kato derived \cite{itoi} the coefficients $A_{S_{T}}$ as
\begin{eqnarray}
	A_{0} = -\frac{4}{3\sqrt{2}}, \hspace{1em} A_{1} = \frac{1}{6\sqrt{2}}, \hspace{1em} A_{2} = \frac{1}{6\sqrt{2}}. \label{ast}
\end{eqnarray}
If $g_{1}(0) \ll -1$, which is corresponding to the ULS model, Eq. \eqref{solg1su3} becomes $g_{1}(l) \approx -2\sqrt{2}/(3\ln N)$.
By substituting this for Eq. \eqref{destg1g2a}, we obtain
\begin{eqnarray}
	\Delta E_{S_{T}} \left(\pm \frac{2\pi}{3} \right) \approx \frac{2\pi v_{0}}{N} \left( x_{S_{T}} - A_{S_{T}} \frac{2\sqrt{2}}{3 \ln N} \right). \label{destg1g2a-app}
\end{eqnarray}
The logarithmic correction in Eq. \eqref{destg1g2a-app} is corresponding to that of the correlation function Eq. \eqref{sqcorre-in}.

\subsection{The case of $g_{1}(l)+g_{2}(l)=0$}
\label{subsec:g1g20}
In the case of $g_{1}(l)+g_{2}(l)=0$, Eqs. \eqref{rg3} and \eqref{rg4} are reduced to
\begin{eqnarray}
        \frac{d g_{1}(l)}{dl} = \frac{1}{2\sqrt{2}} g_{1}^{2}(l), \hspace{2em} \frac{d g_{2}(l)}{dl} = -\frac{1}{2\sqrt{2}} g_{2}^{2}(l).\label{rg4b}
\end{eqnarray}
with the solution
\begin{eqnarray}
	g_{1}(l) = -g_{2}(l) = g_{1}(0) \left[1- \frac{1}{2\sqrt{2}} g_{1}(0)l \right]^{-1}. \label{solg1g2}
\end{eqnarray}
As shown in Fig. \ref{fig:flow}, if $g_{1}(0)<0$, the flow is absorbed into the fixed point corresponding to the SU$(3)_{1}$ WZW model.
On the other hand, if $g_{1}(0)>0$, the flow diverges as $\left(g_{1}, g_{2}\right)\rightarrow (\infty, \infty)$, which is corresponding to the self-dual sine-Gordon (SDSG) model \cite{lecheminant} with massive excitation.
This massive SDSG model is equivalent \cite{lecheminant} to level-2 SO(3) Wess--Zumino--Witten [SO$(3)_{2}$ WZW] model perturbed by a marginal current-current interaction, which is also described by the fermion model of Andrei and Destri \cite{andrei} exactly solvable with the Bethe ansatz.

Also, Eq. \eqref{destg1g2} can be rewritten as
\begin{eqnarray}
        \Delta E_{S_{T}} \left(\pm\frac{2\pi}{3} \right) \approx \frac{2\pi v_{0}}{N} \left[ x_{S_{T}} + C_{S_{T}}g_{1}(l) \right], \label{destg1g2b}
\end{eqnarray}
where $C_{S_{T}} \equiv A_{S_{T}}-B_{S_{T}}$.
By solving equations Eq. \eqref{rg4b}, Itoi and Kato derived \cite{itoi} the coefficients $C_{S_{T}}$ as
\begin{eqnarray}
	C_{0} = -\frac{1}{\sqrt{2}}, \hspace{1em} C_{1} = -\frac{1}{2\sqrt{2}}, \hspace{1em} C_{2} = \frac{1}{2\sqrt{2}}, \label{cst}
\end{eqnarray}
which we interpret as the Clebsch-Gordan coefficients of SO(3) $\times$ SO(3) \cite{lecheminant} (see Sec. \ref{subsec:group}).
If $g_{1}(0) \ll -1$, the TL phase, Eq. \eqref{solg1g2} becomes $g_{1}(l) \approx -2\sqrt{2}/\ln N$.
By substituting this for Eq. \eqref{destg1g2b}, we obtain
\begin{eqnarray}
	\Delta E_{S_{T}} \left(\pm \frac{2\pi}{3} \right) \approx \frac{2\pi v_{0}}{N} \left( x_{S_{T}} - C_{S_{T}} \frac{2\sqrt{2}}{\ln N} \right). \label{destg1g2c-app}
\end{eqnarray}

\subsection{The case of $g_{1}(l)=0$}
\label{subsec:g10}
In the case of $g_{1}(l)=0$, Eqs. \eqref{rg3} and \eqref{rg4} are reduced to
\begin{eqnarray}
        \frac{d g_{2}(l)}{dl} &=& -\frac{3}{2\sqrt{2}} g_{2}^{2}(l).\label{rg4c}
\end{eqnarray}
with the solution
\begin{eqnarray}
	g_{2}(l) = g_{2}(0) \left[1+ \frac{3}{2\sqrt{2}} g_{2}(0)l \right]^{-1}. \label{solg1g2b}
\end{eqnarray}
As shown in Fig. \ref{fig:flow}, if $g_{2}(0)>0$, the flow is absorbed into the fixed point corresponding to the SU$(3)_{1}$ WZW model.
On the other hand, if $g_{2}(0)<0$, the flow diverges as $g_{2}\rightarrow\infty$, which explains the Haldane gap \cite{haldane1,haldane2}.

Also, Eq. \eqref{destg1g2} can be rewritten as
\begin{eqnarray}
        \Delta E_{S_{T}} \left(\pm\frac{2\pi}{3} \right) \approx \frac{2\pi v_{0}}{N} \left[ x_{S_{T}} + B_{S_{T}}g_{2}(l) \right]. \label{destg1g2c}
\end{eqnarray}
By substituting Eqs. \eqref{ast} and \eqref{cst} for $C_{S_{T}} = A_{S_{T}}-B_{S_{T}}$, we obtain
\begin{eqnarray}
        B_{0} = -\frac{1}{3\sqrt{2}}, \hspace{1em} B_{1} = \frac{2}{3\sqrt{2}}, \hspace{1em} B_{2} = -\frac{1}{3\sqrt{2}}. \label{bst}
\end{eqnarray}

If $g_{2}(0) \gg 1$, the TL phase, Eq. \eqref{solg1g2b} becomes $g_{2}(l) \approx 2\sqrt{2}/(3\ln N)$.
By substituting this for Eq. \eqref{destg1g2c}, we obtain
\begin{eqnarray}
	\Delta E_{S_{T}} \left(\pm \frac{2\pi}{3} \right) \approx \frac{2\pi v_{0}}{N} \left( x_{S_{T}} + B_{S_{T}} \frac{2\sqrt{2}}{3 \ln N} \right). \label{destg1g2b-app}
\end{eqnarray}

\section{Phase transition and critical phenomena}
\label{sec:tran}
In this section, we show our numerical results of the critical phenomena around the SU(3) symmetric critical point \cite{mashiko2}.
Here, we make use of the conservation of the magnetization and the translational symmetry for our numerical calculations.
For later discussions, in addition to Eq. \eqref{defdestq}, we define the dispersion curve as
\begin{eqnarray}
\Delta E(q) \equiv E(q) - E_{g}, \label{defdeq}
\end{eqnarray}
where $E(q)$ is the lowest energy at the wave number $q$.

\subsection{TL phase--Trimer phase transition} 
\label{subsec:tlt}
In this subsection, we discuss the phase transition between the TL phase and the Trimer phase in the SU(3) asymmetric case $(\theta \ne \pi/4)$.
We investigate the phase transition by changing $\phi$ and fixing $\theta$ at $\theta=7\pi/18$, which is the parameter in the TL phase of the BLBQ model (see Fig. \ref{fig:phase}).

\subsubsection{dispersion curves}
\label{subsubsec:dis70}

As for the dispersion curves, we obtain the numerical results with $N=9$--$18$. 
Here, we obtain $E_{g} = E(0) = E_{0}(0)$.
There are soft modes at $q=0,\pm2\pi/3$ for all system sizes. 
Considering the fact that soft modes appear at $q=0, \pm 2\pi/3$, one should carry out numerical calculations only in cases where $N$ is a multiple of $3$ in later sections as well.

\begin{figure}[t]
 \begin{center}
  \includegraphics[keepaspectratio,scale=0.255]{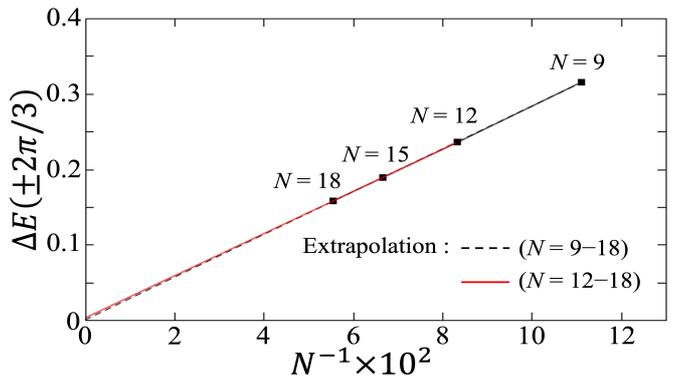}
 \end{center}
        \caption{Elementary excitation energy $\Delta E(\pm 2\pi/3)$ in the case of $\theta=7\pi/18$ and $\phi=0$ as a function of $N^{-1}$. The dashed line and the red solid line show the extrapolation with $N=9$--$18$ and $N=12$--$18$ respectively.}
        \label{fig:den70}
\end{figure}

Figure \ref{fig:den70} illustrates $\Delta E(\pm 2\pi/3) = \Delta E_{0}(\pm 2\pi/3)$ replotted for $N=9$--$18$.
The elementary excitation energy $\Delta E(\pm 2\pi/3)$ depends linearly on $N^{-1}$.
We extrapolate $\Delta E(\pm 2\pi/3)$ with the function $\Delta E(\pm 2\pi/3) = a_{0} + a_{1}N^{-1}$, where $a_{0}$ and $a_{1}$ are constants.
When we extrapolate it with $N=9$--$18$ (dashed line in Fig.  \ref{fig:den70}), we obtain $a_{0}=1.2 \times 10^{-3}$. 
Extrapolating it with $N=12$--$18$ (red solid line in Fig.  \ref{fig:den70}), we obtain $a_{0}=1.4 \times 10^{-3}$.
Although it seems that a small gap remains, this should be massless considering the logarithmic corrections shown in Eqs. \eqref{destg1g2a-app}, \eqref{destg1g2c-app}, and \eqref{destg1g2b-app}.

Lastly, we calculate the spin wave velocity, which we utilize for later calculations of the central charge and scaling dimension. 
Note that the spin wave velocity and the central charge are valid only in the case of the massless TL phase.
The spin wave velocity $v_{0}$ is defined as
\begin{eqnarray}
	v_{0} \equiv \left. \frac{dE(q)}{dq} \right|_{q=0}. \label{v0}
\end{eqnarray}
The velocity is a function of $N$, $v_{0}(N)$. 
In our numerical calculations, we investigate the slope of the dispersion curves to obtain $v_{0}(N)$ written as
\begin{eqnarray}
	v_{0}(N) = \frac{E(2\pi/N) - E(0)}{2\pi/N}. \label{v02}
\end{eqnarray}

\subsubsection{phase boundary}
\label{subsubsec:bound70}
\begin{figure}[t]
 \begin{center}
  \includegraphics[keepaspectratio,scale=0.255]{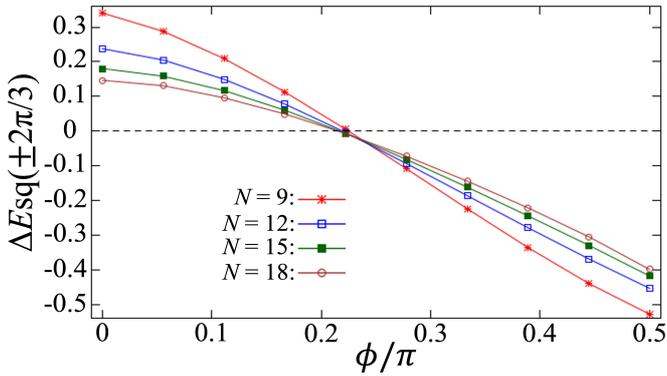}
 \end{center}
        \caption{$\Delta E_{\mathrm{sq}}(\pm 2\pi/3)$ as a function of $\phi$ in the case of $\theta=7\pi/18$ with $N=9$--$18$.}
        \label{fig:bound70}
\end{figure}

Here, we explain how to specify the TL--trimer boundary, shown in Fig. \ref{fig:phase}.
Considering $B_{0}=B_{2}$ in Eq. \eqref{bst}, we plot
\begin{eqnarray}
\Delta E_{\mathrm{sq}}(\pm2\pi/3) \equiv \Delta E_{0} \left(\pm 2\pi/3 \right) - \Delta E_{2} \left(\pm 2\pi/3 \right),
\end{eqnarray}
for various $\phi$ with $N=9$--$18$ in Fig. \ref{fig:bound70} to specify the TL--trimer phase boundary.
As shown in Fig. \ref{fig:bound70}, we firstly find that the value of $\Delta E_{\mathrm{sq}}(\pm 2\pi/3)$ changes from positive to negative at a certain point $\phi_{c}$ for all system sizes.
One can see that the size dependence of the crossing points $\phi_{c}(N)$ is very small.
We discuss these numerical results on the basis of the theory of Itoi and Kato \cite{itoi}. 
\begin{figure}[b]
 \begin{center}
  \includegraphics[keepaspectratio,scale=0.255]{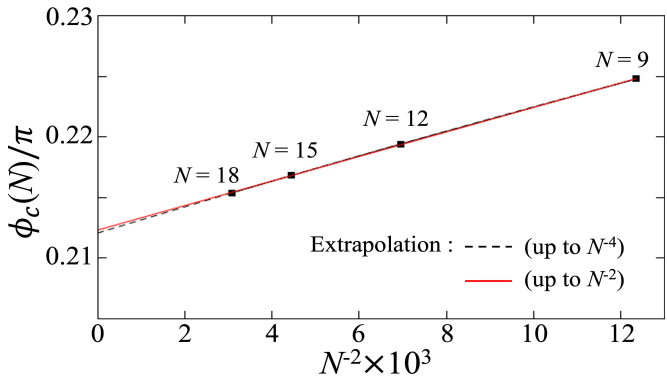}
 \end{center}
        \caption{Phase boundary $\phi_{c}(N)$ as a function of $N^{-2}$ in the case of $\theta=7\pi/18$ with $N=9$--$18$. The dashed line and the red solid line show the extrapolation up to therms of $N^{-4}$ and $N^{-2}$ respectively.}
        \label{fig:phicn70}
\end{figure}

The TL--Trimer phase transition corresponds to the transition between the first quadrant ($g_{1}>0$ and $g_{2}>0$) and the second quadrant ($g_{1}<0$ and $g_{2}>0$) in Fig. \ref{fig:flow}.
It occurs at the boundary of the line $g_{1}=0$ and $g_{2}>0$.
By solving the renormalization-group equation Eq. \eqref{rg4c}, they found \cite{itoi} that if the system is massless TL phase, $\Delta E_{\mathrm{sq}}(\pm 2\pi/3)$ satisfies the relation
\begin{eqnarray}
	\Delta E_{\mathrm{sq}} \left( \pm \frac{2\pi}{3} \right) > 0. \label{esep}
\end{eqnarray}
They also found \cite{itoi} that if the system is the trimer phase, $\Delta E_{\mathrm{sq}}(\pm 2\pi/3)$ satisfies the relation
\begin{eqnarray}
	\Delta E_{\mathrm{sq}} \left( \pm \frac{2\pi}{3} \right) < 0. \label{esen}
\end{eqnarray}
By comparing our numerical results in Fig. \ref{fig:bound70} with the theory \cite{itoi}, we find that the region $\phi<\phi_{c}$ is the $\mathbb{Z}_{3}$ symmetric (massless) phase, and the region $\phi>\phi_{c}$ is the $\mathbb{Z}_{3}$ ordered phase. 
In other words, there occurs a phase transition at $\phi=\phi_{c}$.

In Fig. \ref{fig:phicn70}, we plot $\phi_{c}(N)$, the crossing points of  Fig. \ref{fig:bound70}.
The correction terms $O \left( N^{-2} \right)$ of $\phi_{c}(N)$ can be written as \cite{cardy2,rein,kitazawa}
\begin{eqnarray}
	\phi_{c}(N) = \phi_{c} + X_{1} N^{-2} +X_{2} N^{-4} + O\left(N^{-6} \right), \label{phicn70}
\end{eqnarray}
where the leading term $\phi_{c}$ arises from the SU$(3)_{1}$ WZW model.
The correction term of $N^{-2}$ arises from the descendant fields of the identity operator $\hat{1}$ \cite{cardy2,rein,kitazawa}, that is, $\hat{L}_{-2} \hat{\bar{L}}_{-2} \hat{1}$, $\hat{L}_{-4} \hat{1}$, $\hat{L}_{-2}^{2} \hat{1}$, etc., with the scaling dimension $x=4$.
These descendant fields are not included in Eq. \eqref{asu} of the CFT, but ascribed for the lattice model.
In a similar way, the correction terms of $O \left( N^{-4} \right)$ arise \cite{cardy2,rein,kitazawa}.
As shown in Fig. \ref{fig:phicn70}, we extrapolate the data up to therms of $N^{-4}$ (dashed line), and then we obtain $\phi_{c}/\pi = 0.2120$.
Also, extrapolating it up to therms of $N^{-2}$ (red solid line) we obtain $\phi_{c}/\pi = 0.2123$.
The result of Fig. \ref{fig:phicn70} is shown in Fig. \ref{fig:phase} in the case of $\theta=7\pi/18$.
In the same way, we plot the TL--trimer phase boundary in Fig. \ref{fig:phase} in other cases as well.

\subsubsection{central charge}
\label{subsubsec:c70}
\begin{figure}[b]
 \begin{center}
  \includegraphics[keepaspectratio,scale=0.255]{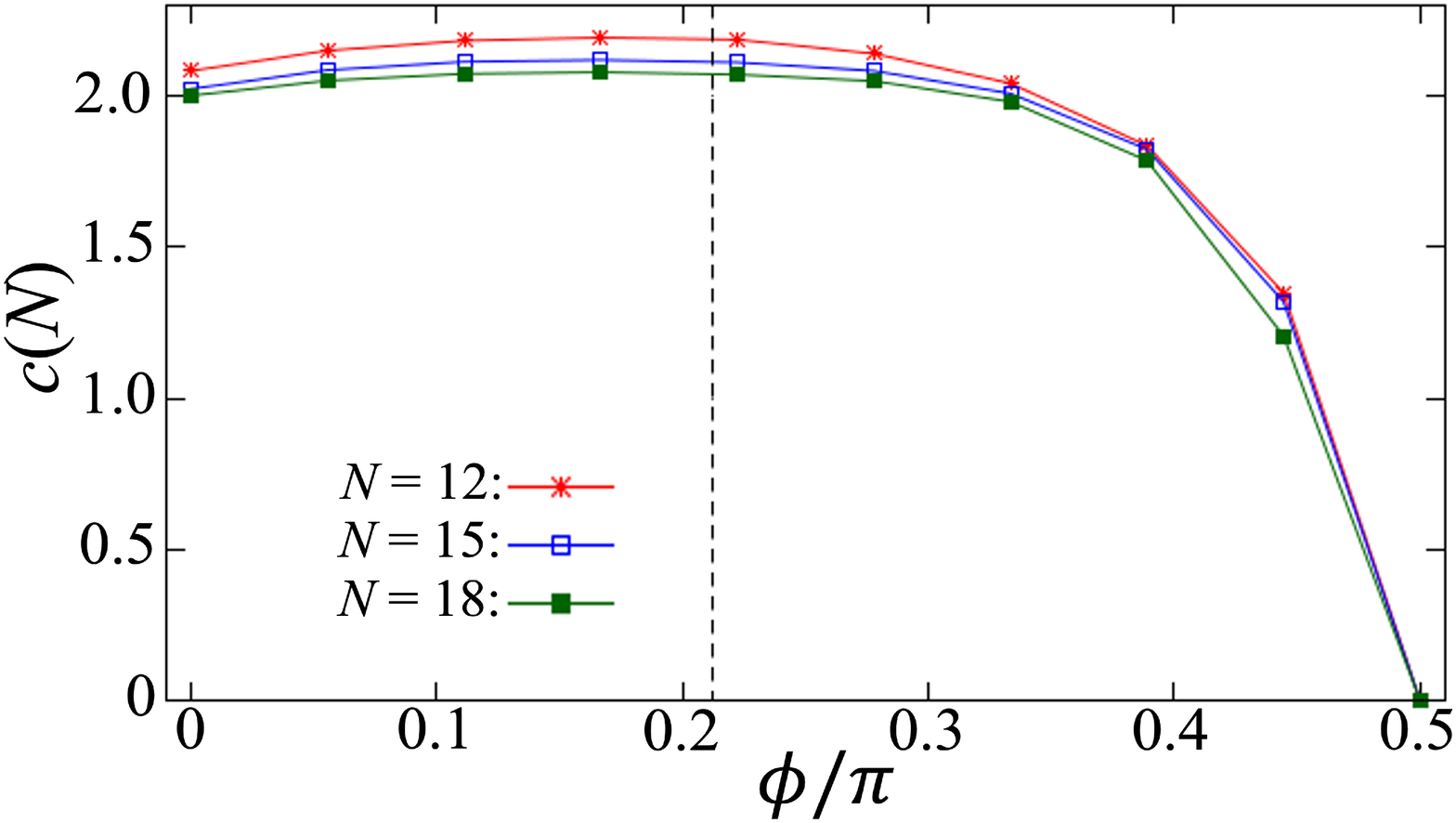}
 \end{center}
	\caption{Effective central charge $c(N)$ as a function of $\phi$ in the case of $\theta=7\pi/18$ with $N=12$--$18$.} 
        \label{fig:c70}
\end{figure}
Here, we calculate the central charge $c$, which characterizes the quantum anomaly and specifies the universality class.
In the critical state of one-dimensional quantum systems under PBCs, the ground-state energy density at $N$ follows \cite{blote,affleck} the equation 
\begin{eqnarray}
	\frac{E_{g}(N)}{N} = \epsilon_{\infty} - \frac{\pi v_{0} c}{6N^{2}}, \label{eg}
\end{eqnarray}
where $\epsilon_{\infty}$ is the ground-state energy density in the case of $N \rightarrow \infty$. 
In Eq. \eqref{eg}, one can numerically calculate $E_{g}$ and $v_{0}$, but there remain two constants, $\epsilon_{\infty}$ and $c$, as unknown values. 
By removing the constant term $\epsilon_{\infty}$ in Eq. \eqref{eg}, we thus calculate the effective central charge $c(N)$ as
\begin{eqnarray}
	&&\frac{E_{g}(N)}{N} - \frac{E_{g}(N-3)}{N-3} \notag \\
	&& \hspace{2em} = -\frac{\pi}{6} \left[\frac{v_{0}(N)}{N^{2}} - \frac{v_{0}(N-3)}{(N-3)^{2}} \right] c(N). \label{eg2}
\end{eqnarray}

\begin{figure}[t]
 \begin{center}
  \includegraphics[keepaspectratio,scale=0.255]{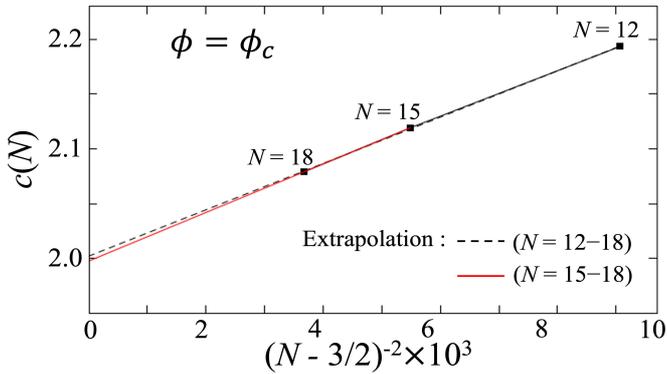}
 \end{center}
	\caption{Effective central charge $c(N)$ as a function of $N^{-2}$ with $N=12$--$18$ at $(\phi,\theta)=(0.21208\pi,7\pi/18)$. The dashed line and the red solid line show the extrapolation with $N=12$--$18$ and $N=15$--$18$ respectively.} 
        \label{fig:cn70}
\end{figure}

Figure \ref{fig:c70} shows the effective central charge as a function of $\phi$ for different system sizes, $N=12$--$18$. 
The effective central charge was firstly numerically calculated in the case of the CFT with $c=1$ \cite{okamoto}. 
In this article, we investigate the effective central charge utilizing Eq. \eqref{eg2}. 
Equation \eqref{eg} is true only in the case of the massless phase certainly, but we can apply Eq. \eqref{eg2} even to systems in a massive phase as well.
We find that the effective central charges smoothly converge to $c=2$ as $N\rightarrow \infty$ in the region $\phi < \phi_{c}$ (Fig. \ref{fig:cn70}).
In Fig. \ref{fig:cn70}, similarly to Eq. \eqref{phicn70}, we extrapolate the effective central charge $c(N)$ as \cite{cardy2,rein,kitazawa}
\begin{eqnarray}
	\hspace{-1em} c(N) &=& c + C_{1} (N-3/2)^{-2} + O\left((N-3/2)^{-4}\right), \label{cf}
\end{eqnarray}
where $C_1$ is  the constants.
We then obtain $c = 2.00$, by extrapolating $c(N)$ with $N=12$--$18$ (dashed line).
Also if we extrapolate it with $N=15$--$18$ (red solid line), we obtain $c = 1.99$.
On the other hand, in the region $\phi > \phi_{c}$, the effective central charge shows a decline as $\phi$ approaches $\pi/2$. 
Moreover, around $\phi=\phi_{c}$, there are no sharp decline of $c(N)$ because of the logarithmic correction \cite{cardy1} $O\left(\left(\ln N \right)^{-3} \right)$.
Comparing our numerical results in Figs. \ref{fig:c70} and \ref{fig:cn70} with Zamolodchikov's $c$-theorem \cite{zamolodchikov} and the theory of Itoi and Kato \cite{itoi}, we find that the region $\phi < \phi_{c}$ is illustrated by the CFT with $c=2$ (critical phase), whereas the region $\phi > \phi_{c}$ is a massive phase.

\subsubsection{scaling dimension}
\label{subsubsec:x70}
Here, we calculate the scaling dimension $x$, which is one of the critical exponents, at the boundary ($\phi=\phi_{c}$). 
We rewrite elementary excitation energy at a certain $S_T$, Eq. \eqref{destg1g2}, as
\begin{eqnarray}
	\Delta E_{S_T} \left( \pm \frac{2\pi}{3} \right)=\frac{2\pi v_{0}}{N}\left[x+A_{S_T}g_{1}(l)+B_{S_T}g_{2}(l) \right]. \label{destg1g2-a}
\end{eqnarray}
In order to remove terms of $g_{1}$ and $g_{2}$ which include logarithmic correction, we calculate the scaling dimension $x$ as 
\begin{eqnarray}
	&& \frac{1}{9} \left[\Delta E_{0} \left( \pm \frac{2\pi}{3} \right) + 3\Delta E_{1} \left( \pm \frac{2\pi}{3} \right) + 5 \Delta E_{2} \left( \pm \frac{2\pi}{3} \right) \right] \notag \\
	&& \hspace{13em} = \frac{2\pi v_{0}(N)}{N}x(N). \label{x}
\end{eqnarray}
We further extrapolate the effective scaling dimension $x(N)$ as \cite{cardy2,rein,kitazawa}
\begin{eqnarray}
        x(N) = x + D_{1} N^{-2} +D_{2} N^{-4} + O\left( N^{-6} \right), \label{xf}
\end{eqnarray}
where $D_{1}$ and $D_{2}$ are constants.
\begin{figure}[t]
 \begin{center}
  \includegraphics[keepaspectratio,scale=0.255]{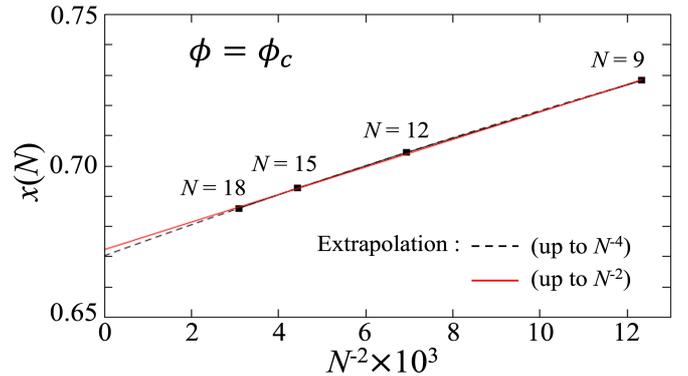}
 \end{center}
        \caption{Effective scaling dimension $x(N)$ with $N=9$--$18$ at $(\phi,\theta)=(0.21208\pi,7\pi/18)$. The dashed line and the red solid line show the extrapolation up to therms of $N^{-4}$ and $N^{-2}$ respectively.} 
        \label{fig:xn70}
\end{figure}
Figure \ref{fig:xn70} shows the effective scaling dimension $x(N)$ at $\phi=\phi_{c}$ with $N=9$--$18$. 
As shown in this figure, we extrapolate $x(N)$ up to therms of $N^{-4}$ (dashed line), and then we obtain $x = 0.670$.
Also, extrapolating it up to therms of $N^{-2}$ (red solid line), we obtain $x = 0.672$.
These numerical results at $\phi=\phi_{c}$ point are in line with the scaling dimension $x=2/3$ of the SU$(3)_{1}$ WZW model \cite{wess,witten1,witten2}.

\subsubsection{other properties}
\label{subsubsec:other70}
\begin{figure}[t]
 \begin{center}
  \includegraphics[keepaspectratio,scale=0.255]{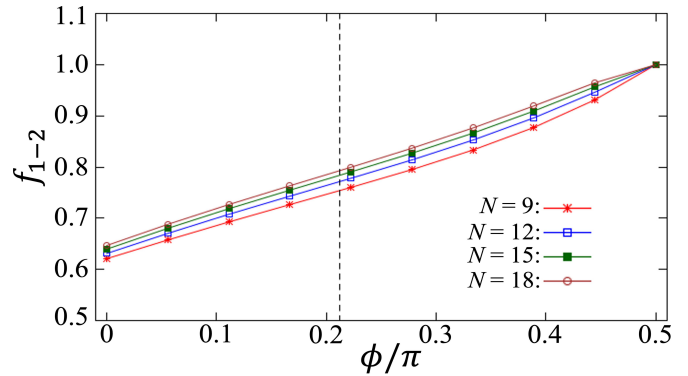}
 \end{center}
	\caption{$f_{1-2}$ with $N=9$--$18$ as a function of $\phi$ at $\theta=7\pi/18$.}
        \label{fig:f1270}
\end{figure}

\begin{figure}[t]
 \begin{center}
  \includegraphics[keepaspectratio,scale=0.255]{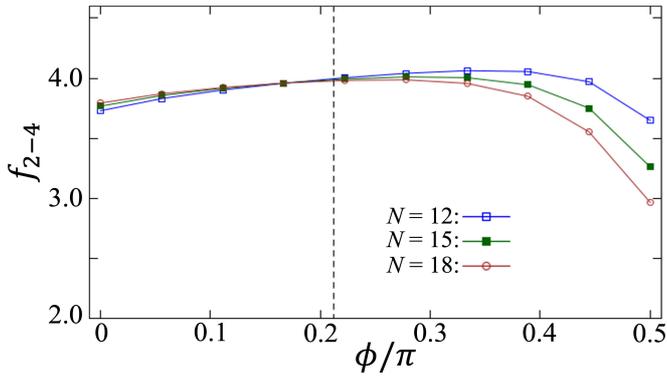}
 \end{center}
        \caption{$f_{2-4}$ with $N=12$--$18$ as a function of $\phi$ at $\theta=7\pi/18$.}
        \label{fig:f2470}
\end{figure}

Here, we confirm the phase transition based on other several properties.
For the discussion, we define $f_{1-2}$ and $f_{2-4}$ as
\begin{eqnarray}
	f_{1-2} \equiv \frac{\Delta E_{2}(\pm 2\pi/3)}{\Delta E_{1}(\pm 2\pi/3)}, \hspace{1em} f_{2-4} \equiv \frac{\Delta E_{4}(\pm 2\pi/3)}{\Delta E_{2}(\pm 2\pi/3)}. \label{deff12andf24}
\end{eqnarray}
We show the numerical results of $f_{1-2}$ and $f_{2-4}$ in Figs. \ref{fig:f1270} and \ref{fig:f2470} respectively.

Firstly, we discuss properties in the vicinity of $\phi=0$, TL phase.
Figure \ref{fig:f1270} shows that $f_{1-2}$ gradually gets larger as $N$ gets larger.
In the case of the thermodynamical limit $N \rightarrow \infty$, it is expected to converge to 1 due to the consistency with the SU(3) symmetric property of the WZW model \cite{wess,witten1,witten2}.
But the convergence is very slow since the excitation energies have the logarithmic corrections \cite{itoi}.
In Fig. \ref{fig:f2470}, $f_{2-4}$ approaches to $4$ as $N$ gets larger, which means the characteristics caused by the spin quadrupole \cite{sutherland}.
Although the convergence is very slow as well, $f_{2-4}$ is nearly $4$ since the effects of the logarithmic corrections is small.

Secondly, we discuss the vicinity of $\phi=\pi/2$, trimer phase.
Figure \ref{fig:f1270} shows $f_{1-2} = 1$ despite the value of $N$ in the case of $\phi=\pi/2$, which means the SU(3) symmetric characteristics described in Eqs. \eqref{ast}.
In Fig. \ref{fig:f2470}, $f_{2-4}$ gradually gets smaller as $N$ gets larger.
We expect that it converge to $2$ as $N \rightarrow \infty$, since two magnons with the magnetization $S^{z}=2$ exist almost independently \cite{greiter}.
If we can deal with larger size, we will observe the convergence to $2$ more sharply.

\subsection{Haldane phase--Trimer phase transition}
\label{subsec:haltri}
In this subsection, we discuss the phase transition between the Haldane phase and the Trimer phase.
We investigate the phase transition by changing $\phi$ and fixing $\theta$ at $\theta=5\pi/36$, which is the parameter in the Haldane phase of the BLBQ model (see Fig. \ref{fig:phase}).
\begin{figure}[t]
 \begin{center}
  \includegraphics[keepaspectratio,scale=0.255]{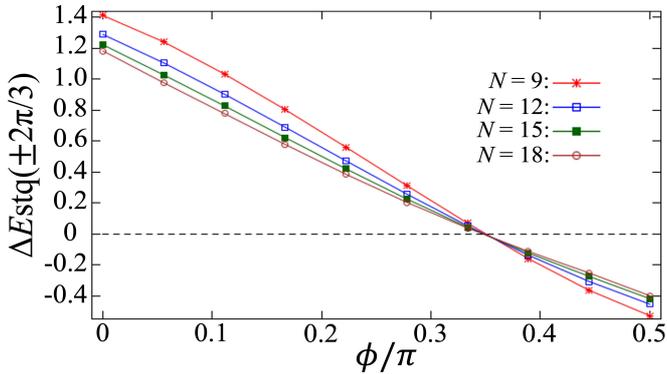}
 \end{center}
        \caption{$\Delta E_{\mathrm{stq}}(\pm 2\pi/3)$ as a function of $\phi$ in the case of $\theta=5\pi/36$ with $N=9$--$18$.}
        \label{fig:bound25}
\end{figure}
\begin{figure}[t]
 \begin{center}
  \includegraphics[keepaspectratio,scale=0.255]{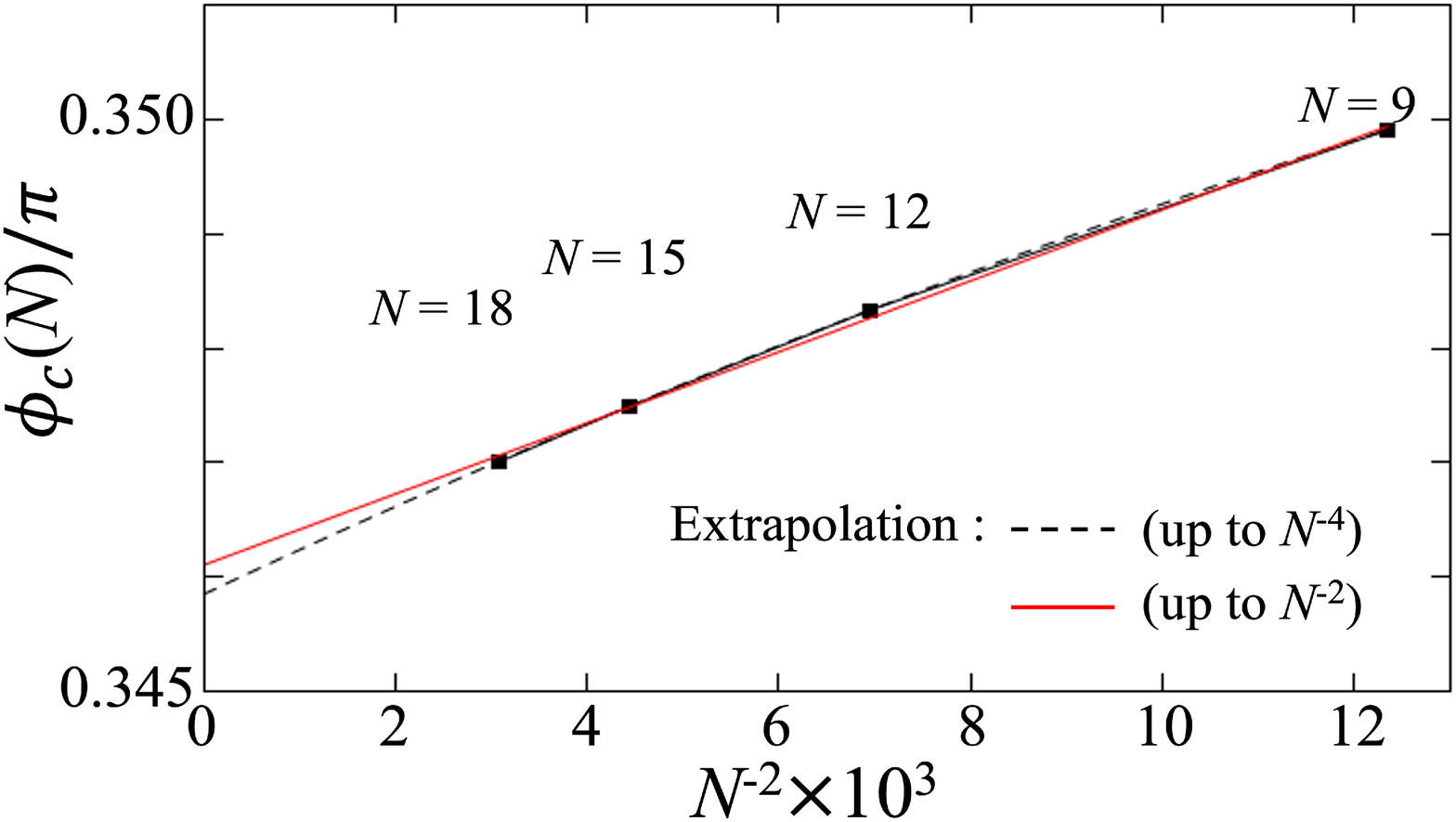}
 \end{center}
        \caption{Phase boundary $\phi_{c}(N)$ as a function of $N^{-2}$ in the case of $\theta=5\pi/36$ with $N=9$--$18$. The dashed line and the red solid line show the extrapolation up to therms of $N^{-4}$ and $N^{-2}$ respectively.}
        \label{fig:phicn25}
\end{figure}

\subsubsection{phase boundary}
\label{subsubsec:bound25}
Here, we explain how to specify the Haldane--trimer boundary, shown in Fig. \ref{fig:phase}.
In Fig. \ref{fig:bound25}, we plot
\begin{eqnarray}
	\Delta E_{\mathrm{stq}}(q) \equiv \Delta E_{0}(q) - \frac{3}{2} \Delta E_{1}(q) + \frac{1}{2} \Delta E_{2}(q), \label{destqq-def}
\end{eqnarray}
for various $\phi$ with $N=9$--$18$, considering the Clebsch-Gordan coefficients of SO(3) $\times$ SO(3) in Eq. \eqref{cst}.
As shown in Fig. \ref{fig:bound25}, we firstly find that the value of $\Delta E_{\mathrm{stq}}(\pm 2\pi/3)$ changes from positive to negative at a certain point $\phi_{c}$ for all system sizes.
One can see that the size dependence of the crossing points $\phi_{c}(N)$ is very small.
We discuss these numerical results on the basis of the theory of Itoi and Kato \cite{itoi}. 

The Haldane--Trimer phase transition corresponds to the transition at the boundary of the line $g_{1}+g_{2}=0$ and $g_{1}>0$) in Fig. \ref{fig:flow}.
By solving the renormalization-group equation Eq. \eqref{rg4c}, they found \cite{itoi} that if the system is Haldane phase, $\Delta E_{\mathrm{stq}}(\pm 2\pi/3)$ satisfies the relation
\begin{eqnarray}
	\Delta E_{\mathrm{stq}} \left( \pm \frac{2\pi}{3} \right) > 0. \label{esetp}
\end{eqnarray}
They also found \cite{itoi} that if the system is the trimer phase, $\Delta E_{\mathrm{stq}}(\pm 2\pi/3)$ satisfies the relation
\begin{eqnarray}
	\Delta E_{\mathrm{stq}} \left( \pm \frac{2\pi}{3} \right) < 0. \label{esetn}
\end{eqnarray}
By comparing our numerical results in Fig. \ref{fig:bound25} with the theory \cite{itoi}, we find that the region $\phi<\phi_{c}$ is the Haldane phase, and the region $\phi>\phi_{c}$ is the $\mathbb{Z}_{3}$ ordered phase. 
In other words, there occurs a phase transition at $\phi=\phi_{c}$.
Also, the boundary, which is the case of $\Delta E_{\mathrm{stq}}(\pm 2\pi/3)=0$, is explained \cite{lecheminant} by SO(3) $\times$ SO(3) SDSG model with massive excitation.

We plot $\phi_{c}(N)$, the crossing point of  Fig. \ref{fig:bound25}, in Fig. \ref{fig:phicn25}.
The correction terms of $\phi_{c}(N)$ behaves \cite{cardy2,rein,kitazawa} as
\begin{eqnarray}
	\phi_{c}(N) = \phi_{c} + Y_{1} N^{-2} +Y_{2} N^{-4} + O\left(N^{-6} \right), \label{phicn25}
\end{eqnarray}
where $Y_{1}$ and $Y_{2}$ are non-universal constants. 
As shown in Fig. \ref{fig:phicn25}, we extrapolate the data up to therms of $N^{-4}$ (dashed line), we obtain $\phi_{c}/\pi = 0.3458$.
Also, extrapolating it up to therms of $N^{-2}$ (red solid line), we obtain $\phi_{c}/\pi = 0.3461$.
The result of Fig. \ref{fig:phicn25} is shown in Fig. \ref{fig:phase} in the case of $\theta=5\pi/36$.
In the same way, we plot the Haldane--trimer phase boundary in Fig. \ref{fig:phase} in other cases as well.
\begin{figure}[b]
 \begin{center}
  \includegraphics[keepaspectratio,scale=0.255]{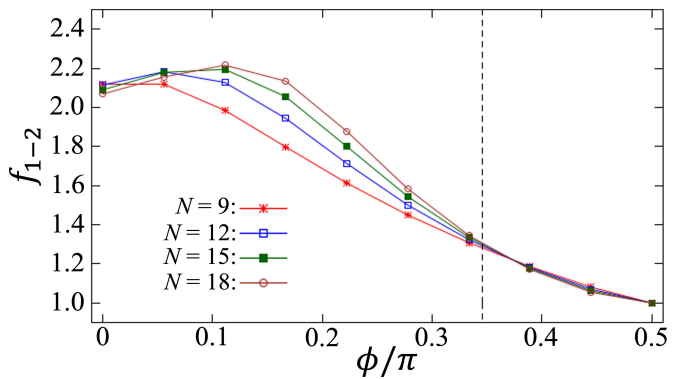}
 \end{center}
        \caption{$f_{1-2}$ with $N=9$--$18$ as a function of $\phi$ in the case of $\theta=5\pi/36$.}
        \label{fig:f1225}
\end{figure}
\begin{figure}[t]
 \begin{center}
  \includegraphics[keepaspectratio,scale=0.255]{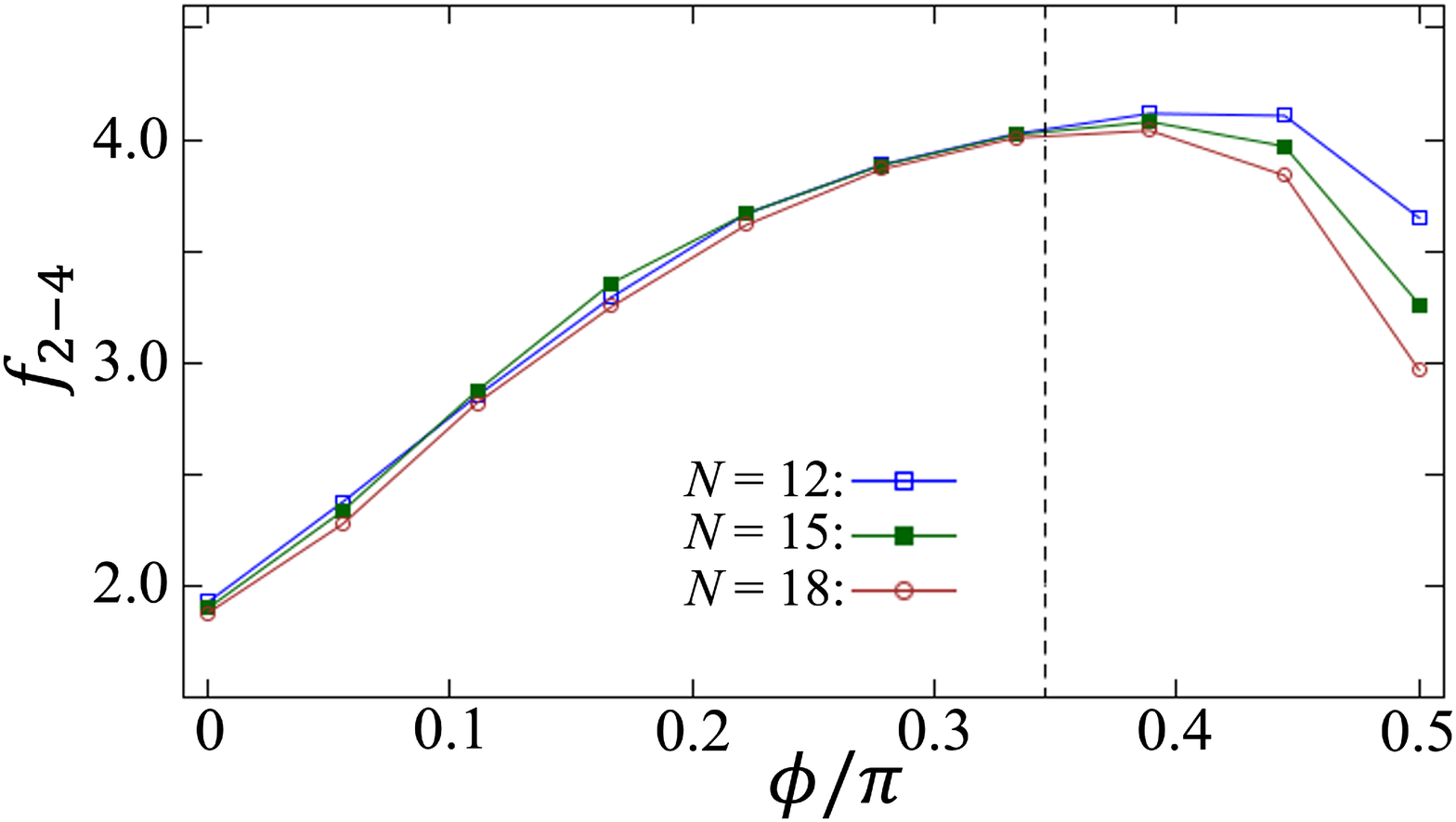}
 \end{center}
        \caption{$f_{2-4}$ with $N=9$--$18$ as a function of $\phi$ in the case of $\theta=5\pi/36$.}
        \label{fig:f2425}
\end{figure}
\begin{figure}[t]
 \begin{center}
  \includegraphics[keepaspectratio,scale=0.255]{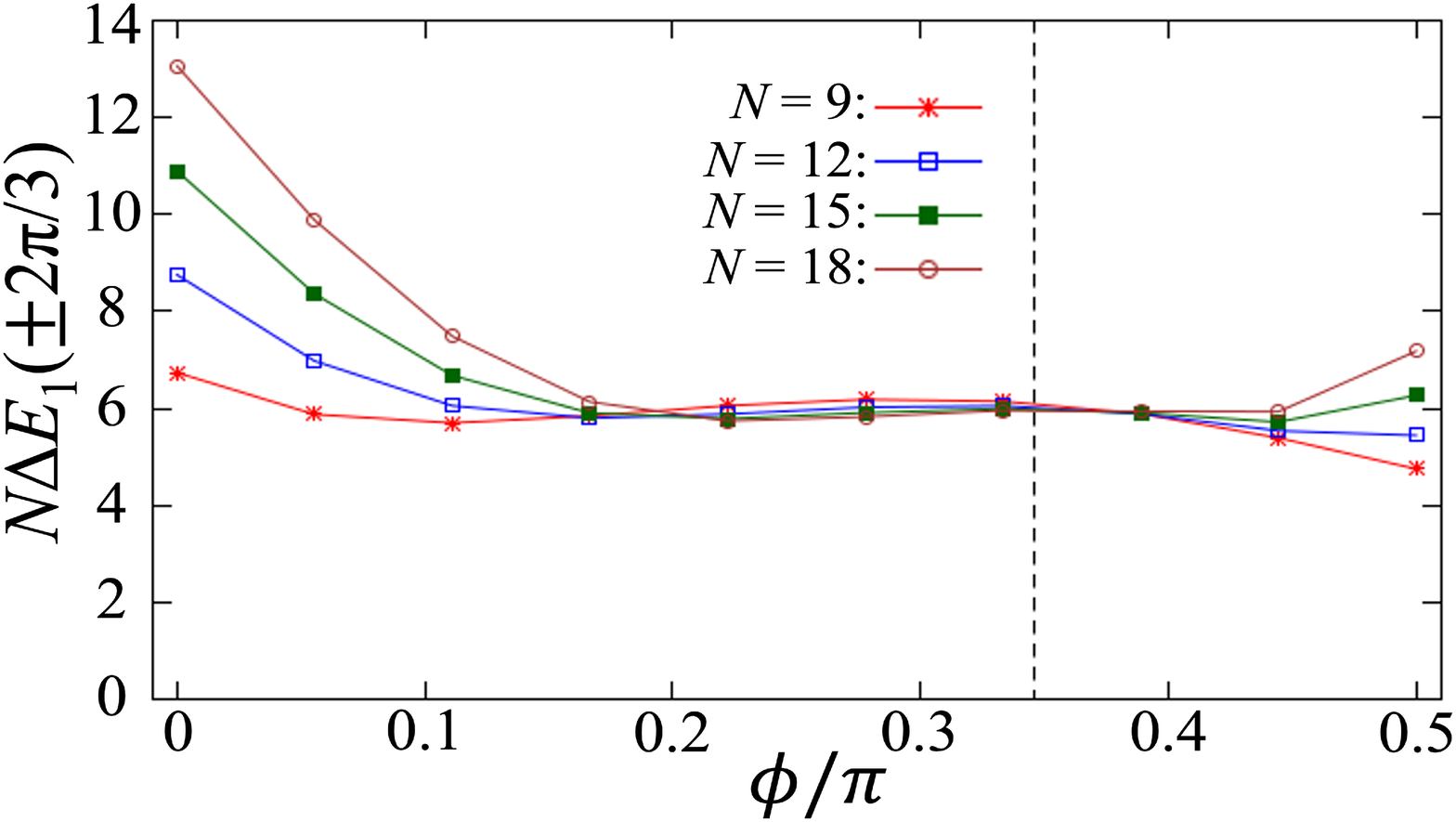}
 \end{center}
	\caption{$N\Delta E_{1}(2\pi/3)$ with $N=9$--$18$ as a function of $\phi$ in the case of $\theta=5\pi/36$.}
        \label{fig:nde125}
\end{figure}
\begin{figure}[t]
 \begin{center}
  \includegraphics[keepaspectratio,scale=0.255]{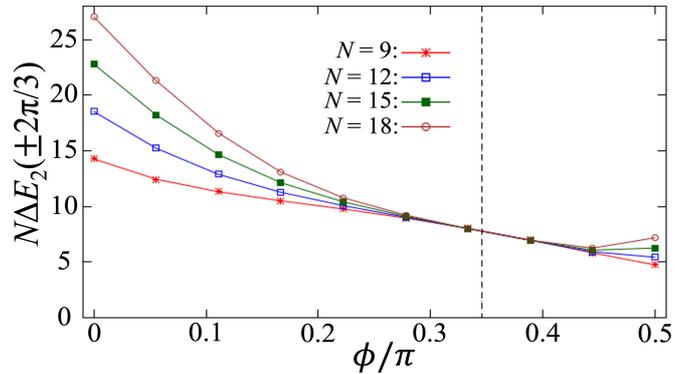}
 \end{center}
        \caption{$N\Delta E_{2}(2\pi/3)$ with $N=9$--$18$ as a function of $\phi$ in the case of $\theta=5\pi/36$.}
        \label{fig:nde225}
\end{figure}

\subsubsection{other properties}
\label{subsubsec:other25}
Here, we confirm the phase transition based on other several properties.
We show the numerical results of $f_{1-2}$ and $f_{2-4}$ in Figs. \ref{fig:f1225} and \ref{fig:f2425} respectively.

Firstly, we discuss properties in the vicinity of $\phi=0$, Haldane phase.
In Figs. \ref{fig:f1225} and \ref{fig:f2425}, $f_{1-2}$ and $f_{2-4}$ approach to $2$ as $N$ gets larger.
This is consistent with the property illustrated with the nonlinear sigma model \cite{haldane1,haldane2}, in which there are two massive magnons independently.

Secondly, Figs. \ref{fig:nde125} and \ref{fig:nde225} show the numerical results of $N \Delta E_{1}(2\pi/3)$ and $N \Delta E_{2}(2\pi/3)$ respectively.
Both figures show behaviors of second-ordered transition, although the Haldane--Trimer phase transition should be the first-ordered transition \cite{lecheminant} illustrated with the massive SO(3) $\times$ SO(3) SDSG model.
We believe that this is because we investigate the phase transition near the fixed point $(g_{1},g_{2})=(0,0)$, which shows the critical state described by the SU$(3)_{1}$ WZW model \cite{wess,witten1,witten2}.
If we investigate it in the region far from the fixed point, we will see the behavior of the first-ordered transition, which is the future work.
Note that the analysis based on the CFT may be invalid to investigate properties of the first-order transition in principle.
In any case, the elucidation of the first-order transition is one of the future tasks.


\subsection{TL phase--Haldane phase transition} 
\label{subsec:haltl}
As for the TL phase--Haldane phase transition, it is exactly shown \cite{itoi} that SU(3) symmetric line ($g_{2}=0$, $g_{1}<0$ in Fig. \ref{fig:flow}) is the phase boundary.
Also, the universality class is the BKT-like with $c=2$ and $x=2/3$ \cite{itoi}.
There are several numerical calculations \cite{fath,schm,lauch,oh,mashiko1} on this phase transition.
Therefore, we do not do the numerical calculation in this paper.

\section{CONCLUSION AND DISCUSSION}
\label{sec:con}
\subsection{Summary}
We have investigated the model Eq. \eqref{model} to reveal the critical phenomena when changing two parameters $\theta$ and $\phi$.
First of all, we find the tri-critical point among three phases, the TL phase, the trimer phase, and the Haldane phase, which corresponds with the SU(3) symmetric critical point found in Ref. \cite{mashiko2} described with the SU$(3)_{1}$ WZW model \cite{wess,witten1,witten2}.
We specify the tri-critical point is $(\theta, \phi) = (\pi/4, 0.22339\pi)$ and the phase diagram is shown in Fig. \ref{fig:phase}. 
Secondly, the critical phenomena belong to the Berezinskii--Kosterlitz--Thouless (BKT)-like universality class with the central charge $c=2$ and the scaling dimension $x=2/3$.
Thirdly, the boundary between the Haldane phase and the trimer phase, $g_{1}+g_{2}=0$ and $g_{1}>0$ in Fig. \ref{fig:flow}, is illustrated with the SDSG model \cite{lecheminant} of massive spin current.

\subsection{Discussion based on the group theory}
\label{subsec:group}
Here, we discuss the degeneracy shown in Secs. \ref{subsec:g20}--\ref{subsec:g10} based on the representation of group theory.
First of all, in the case of the line $g_{2}=0$, there is the degeneracy of triplet and quintuplet states as shown in Eq. \eqref{ast}, which is corresponding to the direct product
\begin{eqnarray}
\bm{3}_{\mathrm{SU(3)}} \otimes \bar{\bm{3}}_{\mathrm{SU(3)}} = \bm{1}_{\mathrm{SU(3)}}  \oplus \bm{8}_{\mathrm{SU(3)}}, \label{directpro3b318}
\end{eqnarray}
where $\bm{3}_{\mathrm{SU(3)}}$, $\bar{\bm{3}}_{\mathrm{SU(3)}}$, $\bm{1}_{\mathrm{SU(3)}}$, and $\bm{8}_{\mathrm{SU(3)}}$ are representations of SU(3).
Eq. \eqref{directpro3b318} reflects the characteristics of the SU(3) symmetric case of models Eqs. \eqref{model} and \eqref{asu}.

Secondly, the line $g_{1}+g_{2}=0$ is explained with the group SO(3), that is, the subgroup of SU(3).
In this case, the model can be represented \cite{lecheminant} by the representation of SO(3) $\times$ SO(3) as
\begin{eqnarray}
\bm{3}_{\mathrm{SO(3)}} \otimes \bm{3}_{\mathrm{SO(3)}} = \bm{1}_{\mathrm{SO(3)}}  \oplus \bm{3}_{\mathrm{SO(3)}} \oplus \bm{5}_{\mathrm{SO(3)}}, \label{directpro33135}
\end{eqnarray}
where $\bm{3}_{\mathrm{SO(3)}}$, $\bm{1}_{\mathrm{SO(3)}}$, $\bm{3}_{\mathrm{SO(3)}}$, and $\bm{5}_{\mathrm{SO(3)}}$ are representations of SO(3).
By comparing Eqs. \eqref{cst}and \eqref{directpro33135} with Ref. \cite{lecheminant}, we interpret Eq. \eqref{cst} as the Clebsch-Gordan coefficients of SO(3) $\times$ SO(3).

Lastly, in the case of the line $g_{1}=0$, we obtain the degeneracy of singlet and quintuplet states as shown in Eq. \eqref{bst}.
We expect this degeneracy to correspond to the relation 
\begin{eqnarray}
\bm{3}_{\mathrm{SU(3)}} \otimes \bm{3}_{\mathrm{SU(3)}} = \bar{\bm{3}}_{\mathrm{SU(3)}}  \oplus \bm{6}_{\mathrm{SU(3)}}, \label{directpro33b36}
\end{eqnarray}
where $\bm{6}_{\mathrm{SU(3)}}$ is the representation.

Therefore, we consider that the perturbative renormalization group of Itoi and Kato \cite{itoi} may be supported with the symmetry, even in the case far away from the fixed point $g_{1}=g_{2}=0$.

\subsection{Correlation functions}
\label{subsec:corre}
Here, we discuss several correlation functions in each phase to discuss the critical behavior.
As for the TL phase and the trimer phase, we already discussed \cite{mashiko2} them based on Refs. \cite{itoi,stou,lauch,schm}, utilizing the spin order parameter $\hat{\bm{S}}_{i}$, the spin-quadrupolar order parameter $\hat{Q}^{\mu\nu}_{(i)}$, and the order parameter of the trimer $\hat{T}_{i}$ defined as
\begin{eqnarray}
	&& \hspace{-2em} \hat{Q}^{\mu\nu}_{(i)} \equiv \frac{1}{2} \left\{\hat{S}^{\mu}_{i}, \hat{S}^{\nu}_{i} \right\} - \frac{2}{3}\delta^{\mu\nu}, \label{sqorder} \\
	&& \hspace{-2em} \hat{T}_{i} \equiv \hat{T}^{\mathrm{P}}_{i} - \left \langle \hat{T}^{\mathrm{P}}_{i} \right \rangle, \hspace{1em} \hat{T}^{\mathrm{P}}_{i} \equiv \left| \{i,j,k \} \right \rangle \left \langle \{i,j,k \} \right|, \label{ordertri}
\end{eqnarray}
where the state vector of $\left| \{i,j,k \} \right \rangle$ is the same as $\{ \circ \circ \circ \}$, Eq. \eqref{trimer2}, and we put $j\equiv i+1$ and $k\equiv i+2$.
As for the TL phase, the correlation functions are expected \cite{itoi,stou,lauch} to be
\begin{eqnarray}
	 \left \langle \hat{\bm{S}}_{i} \cdot \hat{\bm{S}}_{i+r}  \right \rangle &\propto& \cos \left(\frac{2\pi}{3}r \right) r^{-4/3} \left(\ln r \right)^{\sigma_{1}}, \label{scorre} \\
	 \left \langle \hat{Q}^{\mu\nu}_{(i)} \hat{Q}_{(i+r)\mu\nu}  \right \rangle &\propto& \cos \left(\frac{2\pi}{3}r \right) r^{-4/3} \left(\ln r \right)^{\sigma_{2}}, \label{qcorre} \\
	 \left \langle \hat{T}_{i} \hat{T}_{i+r} \right \rangle &\propto& \cos \left(\frac{2\pi}{3}r \right) r^{-4/3} \left(\ln r \right)^{\sigma_{3}}, \label{tprocorre}
\end{eqnarray}
from $c=2$ and $x=2/3$.
Here, $\sigma_{a}$ ($a=1,2,3$) are critical exponents of the logarithmic correction.
Especially, in the SU(3) symmetric case ($g_{2}=0$), critical exponents are $\sigma_{1}=\sigma_{2}=2/9$ and $\sigma_{3}=-16/9$ \cite{itoi}.
In the non-SU(3) case, they take different values.

Secondly, in the trimer phase, the correlation functions are expected \cite{itoi,schm} to be
\begin{eqnarray}
	 \hspace{0em} \left \langle \hat{\bm{S}}_{i} \cdot \hat{\bm{S}}_{i+r}  \right \rangle &\propto& \cos \left(\frac{2\pi}{3}r \right) e^{-r/\xi}, \label{scorrez3} \\
	 \hspace{0em} \left \langle \hat{Q}^{\mu\nu}_{(i)} \hat{Q}_{(i+r)\mu\nu}  \right \rangle &\propto& \cos \left(\frac{2\pi}{3}r \right) e^{-r/\xi}, \label{qcorrez3} \\
	 \hspace{0em} \left \langle \hat{T}_{i} \hat{T}_{i+r} \right \rangle &\propto& \cos \left(\frac{2\pi}{3}r \right), \label{tprocorrez3}
\end{eqnarray}
where $\xi$ is the correlation length.
Equation \eqref{tprocorrez3} characterizes a trimer long-range order.
The Haldane phase is characterized by the correlation function \cite{haldane1,haldane2,nomura3} as
\begin{eqnarray}
\left \langle \hat{\bm{S}}_{i} \cdot \hat{\bm{S}}_{i+r}  \right \rangle \propto \cos \left(\pi r \right) r^{-1/2} e^{-r/ \xi}, \label{sqcorrez4}
\end{eqnarray}
with short-range correlation.

\subsection{Future works and applications}
\label{subsec:open}

There remains an unsolved problem as for the Haldane--trimer transition.
In our numerical calculations, we could not verify that this transition is the first-ordered transition, which Lecheminant and Totsuka expected \cite{lecheminant}.
We believe that the boundary line in Fig. \ref{fig:phase} determined with the properties of SO(3) $\times$ SO(3) is reasonable, although there remain unsolved problems as for the universality class of this phase transition.
Also, we should investigate the incommensurability of the dispersion curves in order to discuss the properties of the sub-phases in the Haldane phase and trimer phase \cite{nomura2,murashima}.


As  for applications, we believe that our theories and results are applicable to experiments of ultracold atomic systems in an optical lattice, described by the SU($\nu$) ($\nu$: integer) Hubbard model \cite{hubbard}, for example, SU(6) symmetric $^{173}$Yb atomic systems \cite{desalvo} or SU(10) symmetric $^{87}$Sr atomic systems \cite{gorshkov}.
Also, Greiter and Rachel expected \cite{greiter} the existence of the atomic systems of the SU(3) spins in an optical lattice with internal interaction composing the trimer state.

\section{acknowledgments}
We also appreciate the advice from K. Totsuka who gave us as for the relationship between the Itoi--Kato model and the self-dual sine-Gordon model. 
T. Mashiko is financially supported by the grant of Japan Science and Technology Agency, No. JPMJSP2136.
K. Nomura is supported by Japan Society for the Promotion of Science KAKENHI Grants, No. 21H05021.


%
\end{document}